# Scalable Fairness Shaping with LLM-Guided Multi-Agent Reinforcement Learning for Peer-to-Peer Electricity Markets

Shrenik Jadhav[1], Birva Sevak[1], Srijita Das[1], Akhtar Hussain[2], Wencong Su[3,*], Van-Hai Bui[3,*]

[1]*Departmernt of Computer and Information Science, University of Michigan-Dearborn, USA*

[2]*Departmernt of Electrical and Computer Engineering, Laval University, Canada*

[3]*Departmernt of Electrical and Computer Engineering, University of Michigan-Dearborn, USA*

**Abstract**

Peer-to-peer (P2P) energy trading is becoming central to modern distribution systems as rooftop PV and home energy management systems become pervasive, yet most existing market and reinforcement learning designs emphasize efficiency or private profit and offer little real-time guidance to ensure equitable outcomes under uncertainty. To address this gap, a fairness-aware multiagent reinforcement learning framework, FairMarket-RL, is proposed in which a large language model (LLM) critic shapes bidding policies within a continuous double auction under partial observability and discrete price–quantity actions. After each trading slot, the LLM returns normalized fairness scores Fairness-to-Grid (FTG), Fairness-Between-Sellers (FBS), and Fairness-of-Pricing (FPP) that are integrated into the reward via ramped coefficients and tunable scaling, so that fairness guidance complements, rather than overwhelms, economic incentives. The environment models realistic residential load and PV profiles and enforce hard constraints on prices, physical feasibility, and policy-update stability. Across a progression of experiments from a small pilot to a larger simulated community and a mixed-asset real-world dataset, the framework shifts exchanges toward local P2P trades, lowers consumer costs relative to grid-only procurement, sustains strong fairness across participants, and preserves utility viability. Sensitivity analyses over solar availability and aggregate demand further indicate robust performance, suggesting a scalable, LLM-guided pathway to decentralized electricity markets that are economically efficient, socially equitable, and technically sound.

**Keywords:** Continuous double auction, fairness shaping, large language models, multi-agent reinforcement learning, peer-to-peer energy trading.

## 1. Introduction

The rapid integration of distributed energy resources (DERs) at the grid edge rooftop photovoltaic arrays, behind-the-meter batteries, and smart meters has turned formerly passive customers into ‘prosumers’ who can generate, store, and trade electricity locally [1]. This technological shift has motivated the design of peer-to-peer (P2P) electricity markets, in which households exchange surplus energy directly rather than routing every kilowatt-hour through a central utility [2].

Reported benefits include reduced network losses, deferred distribution upgrades, and improved consumer choice [3]. Implementing a self-governing P2P market, however, remains challenging. Individual households are small, intermittent producers whose private profit motives do not automatically align with community goals such as grid independence or equitable revenue sharing [4]. Conventional rule-based tariffs handle neither this behavioral diversity nor the stochasticity of renewable generation, while purely profit-driven bidding strategies can lead to market domination by early adopters or excessive reliance on the main grid [5]. Recent work proposes reinforcement learning (RL) agents for local trading, yet most studies optimize private reward functions without explicit fairness guarantees [6]. To provide a structured view of this interaction landscape, Figure 1 illustrates the relationship between participating agents prosumers, consumers, the grid, and the P2P market environment within which they trade.

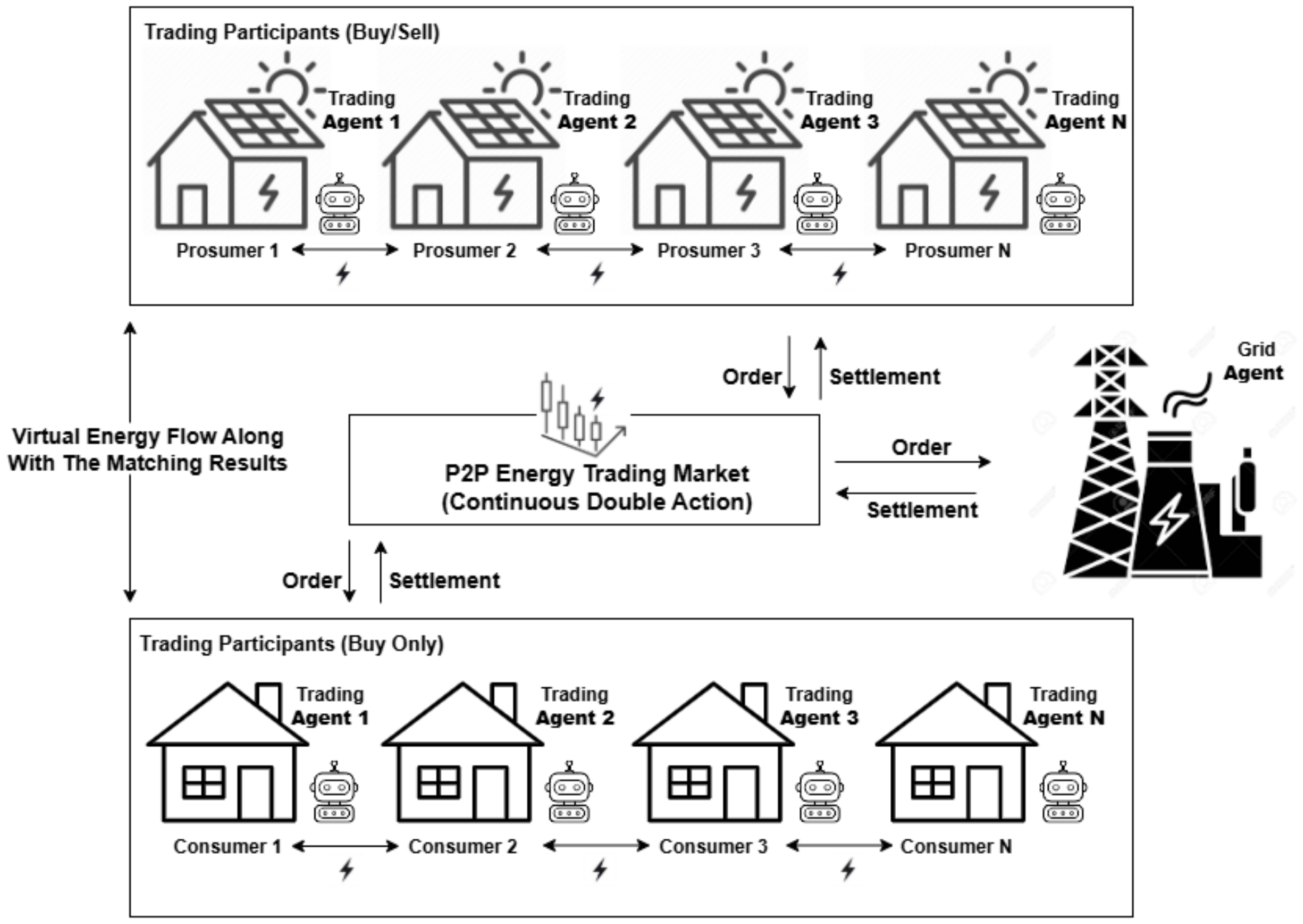

**Figure 1**. Overview of agent interactions in a P2P energy trading system.

Recent developments have seen reinforcement learning, and multi-agent reinforcement learning (MARL), applied to decentralized P2P energy markets, enabling agents to learn adaptive bidding strategies without relying on explicit system models [7], [8]. For instance, actor–critic methods have been deployed to optimize sequential prosumer interactions within continuous market environments [9], while federated learning architectures support scalable training across distributed agents without central data pooling [10]. These RL-based approaches have demonstrated promise in reducing grid reliance and improving overall trading efficiency across various case studies [7], [11]. However, most frameworks optimize either individual payoff or global cost and lack mechanisms to enforce fairness constraints, such as seller equality, price dispersion, or grid-dependency limits [12], [13]. Additionally, while some recent works incorporate physical power-system or operational constraints into RL formulations [14], none provide direct, real-time fairness feedback during training, leaving open the risk of imbalanced trades or systemic reliance on the central grid even when economic metrics are met.

Traditional approaches to P2P energy trading have often relied on static optimization and rule-based clearing with fixed pricing contracts or ex-ante bidding rules, which lack adaptability to dynamic, uncertain environments. Lin et al. [15] compared auction mechanisms and bidding strategies for P2P markets, underscoring how mechanism design choices affect fairness and efficiency. However, such mechanisms are often rigid, assuming either truthful bidding or ideal behavior, and fail to account for the heterogeneity in household preferences, production variability, or incomplete information. Subsequent studies introduced agent-based modeling frameworks to simulate P2P interactions under more realistic conditions. For example, Zhou et al. [16] presented a framework in which prosumers and consumers optimally bid based on local forecasts and budget constraints, yet fairness was imposed exogenously via mechanism parameters such as caps or subsidies rather than being integrated into agents' learning objectives. Game-theoretic approaches ranging from Stackelberg formulations for virtual microgrids to broader surveys of market mechanisms have been explored to characterize equilibrium prices and encourage cooperation under uncertainty [17], [18], but they typically presume full observability and perfectly rational behavior, and their performance degrades in decentralized or partially observable settings. Early reinforcement-learning applications to energy trading further limited coordination potential by focusing exclusively on self-interested reward maximization without considering social welfare or fairness constraints. While these foundational efforts were instrumental in establishing the feasibility of P2P trading models, they lacked integrated mechanisms for fairness, grid-impact awareness, and responsiveness to real-time market imbalance.

A wide range of empirical studies now sheds light on both the achievements and ongoing limitations of advanced P2P energy-trading systems across simulated and practical contexts. Qiu et al. [19] evaluated an eight-agent P2P double-auction community using Ausgrid data and showed that their DA-MADDPG algorithm outperformed zero-intelligence and independent RL baselines by lowering total energy bills and increasing internal trades, thereby reducing external grid reliance, though training relied on public order-book information and the environment omitted network constraints. Shah et al. [20] evaluated a dairy-farm P2P trading simulator (MAPDES) and showed that adding a Q-learning agent reduced electricity costs by about 43%, cut peak grid demand by about 42%, and increased energy-sale revenue by about 1.91% relative to a baseline; however, the setup's limited adaptivity still leaves questions about robustness in more dynamic settings. In the domain of fairness, Behrunani et al. [21] propose a hierarchical price-mediation scheme that equalizes the normalized cost reductions across hubs demonstrated on a 3-hub case, achieving network-wide fairness while relying on a semi-decentralized mediator layer that may limit scalability in larger deployments. Ruiz Irusta and Morales [22] applied distributional fairness optimization to a 1,600-agent system, reducing group inequity by 70.1% and enabling full parity when community PV resources were included; however, the method presumes accurate group classifications and the availability of a non-profit community PV resource modeled as zero-price energy to fully eliminate unfairness. Finally, Kusatake et al. [23] used multi-objective evolutionary optimization to trace the Pareto frontier between efficiency (total prosumer benefit) and fairness (standard deviation of benefits) in a 10-prosumer microgrid, showing the expected trade-offs but relying on a centralized, small-scale setup that has yet to be validated for larger, decentralized markets.

However, while substantial progress has been made, persistent gaps remain in the areas of agent coordination, adaptive response to market dynamics, and real-time integration of fairness objectives. Nazari-Heris et al. [26] applied MARL to optimal energy trading and scheduling in integrated electricity-and-heating networks, embedding operational and multi-energy coupling constraints within the learning-and-control setting. At the same time, federated learning has been explored for coordinating multiple smart homes while preserving data privacy [24], [25]. Yet these advances primarily focus on efficiency and constraint satisfaction and stop short of embedding fairness directly into the learning loop. Some studies have begun exploring fairness quantification and equity-aware mechanisms in trading outcomes [21]– [23], [27]. Gupta et al. [27] employed constrained Markov games to balance grid load while encouraging fairness-aware bidding. Yet despite these efforts, most frameworks treat fairness as a post-hoc evaluation criterion rather than a shaping force during policy learning. Moreover, no unified architecture has emerged that enables agents to adaptively respond to real-time fairness signals such as seller imbalance, price dispersion, or grid dependency while maintaining profitability. This reveals a critical design gap: existing MARL-based P2P systems lack real-time, fairness-aware feedback integration that could bridge the divide between local optimization and socially desirable trading behavior.

This study builds upon our earlier work, published as a short letter, FairMarket-RL: LLM-Guided Fairness Shaping for Multi-Agent Reinforcement Learning in Peer-to-Peer Markets [28], which introduced a novel fairness-aware MARL framework for small-scale P2P energy trading. In the present work, we extend this framework to address the previously identified limitations by scaling to larger and more diverse communities, incorporating extended trading horizons, and enhancing robustness under varying operational conditions. Departing from conventional models that focus solely on profit maximization or grid-wide cost minimization, FairMarket-RL introduces a multi-objective reward design that combines traditional economic incentives with live, ledger-based fairness feedback. Specifically, the framework tracks and incorporates three key fairness metrics, Fairness-to-Grid (FTG) to discourage over-reliance on centralized supply, Fairness-by-Share (FBS) to ensure equitable trading opportunities across prosumers, and Fairness-by-Price (FPP) to minimize price discrimination and volatility. These metrics are computed dynamically during each auction slot and fed back into the agents' learning signals in real time. The architecture supports a continuous double auction (CDA) mechanism, where agents interact sequentially in a partially observable, multi-agent environment using Proximal Policy Optimization (PPO). FairMarket-RL is designed to be modular and extensible, accommodating various policy interventions, market structures, and agent configurations without sacrificing scalability. To evaluate its effectiveness, we conduct a detailed case study spanning a 72-hour trading horizon involving multiple prosumers, consumers, and a grid agent. Our results demonstrate that FairMarket-RL consistently improves trading equity, market-participation diversity, and systemic resilience, while maintaining competitive performance in terms of total welfare and trading volume. By explicitly aligning individual agent incentives with community-oriented fairness objectives, FairMarket-RL offers a principled and operational framework for building next-generation, decentralized electricity markets that are not only economically efficient but also socially responsible and technically robust.

## 2. Peer-to-Peer Electricity Market – Problem Statement

Electricity prosumers increasingly sell excess rooftop-PV generation directly to their neighbors, creating neighborhood-scale P2P markets [29]-[32]. We model a community with m prosumers ($agents\ 1, \ldots, m$) and $n$ pure consumers (agents $m + 1, \ldots, m + n$), together with a single price-taking grid. A learning episode spans D consecutive days, yielding $H = 24D$ hourly decision slots indexed by $t$. Each slot is identified by hour-of-day $h_t \in \{0, \ldots, 23\}$ and day index $d_t \in \{0, \ldots, D - 1\}$. A public daily weather flag $w_{\{d_t\}} \in \{1,0\}$ (sunny, cloudy) modulates load and PV forecasts via an intensity factor:

$$\kappa_{d_t} = \alpha + (1 - \alpha) \cdot w_{d_t}, \quad \alpha \in (0,1) \tag{1}$$

Hourly market chronology. Each slot proceeds in six steps:

- Forecasts: Each household receives one-step-ahead forecasts of its own load and, for prosumers, its PV output; forecasts are conditioned on $(w_{d_t}, h_t, d_t)$.
- Prosumers submit requests: Prosumers announce discrete price–quantity offers sequentially.
- Consumers submit bids: Consumers submit purchase-quantity bids at the grid tariff simultaneously.
- Auction clearing (CDA): A continuous double auction matches the lowest asks with the highest bids greedily while the trade condition holds:

  $$\boldsymbol{p_{buy}} \geq \boldsymbol{p_{sell}} \tag{2}$$

  Executing the minimum available quantity at each match and removing/resizing orders until no executable pair remains.
- Grid settlement: Any unmet demand is served by the grid at the retail tariff; any unsold surplus is purchased by the grid at the feed-in tariff.
- Rewards: Agents receive raw economic rewards (prosumers: revenue minus expenditure; consumers: negative expenditure). A large-language model then computes slot-level fairness metrics, which are incorporated as reward shaping to align private profit with community-level fairness.

The decision problem is partially observable: each household observes only its own forecasts and the public signals $(w_{d_t}, h_t, d_t)$.The order book belongs to the simulator's global state but is not observed by agents [33], [34].

Formal game definition:

$$G = \left\langle S, \{A^{s_i}\}_{i=1}^{m}, \{A^{b_j}\}_{j=m+1}^{m+n}, T, R \right\rangle \tag{3}$$

• State S. At slot $t$, S includes true loads $L_{a,t}$, PV outputs $G_{a,t}$ , the weather flag $w_{d_t}$, time indices $h_t, d_t$, and the current order book; for any storage-enabled prosumer, $S$ also includes battery state-of-charge $B_a, t$.

• Actions. Prosumers choose a discrete price–quantity ask $(p, q)$; storage-enabled prosumers additionally choose charge/discharge/idle decisions (and amounts) subject to power and SOC limits. Consumers choose purchase quantities at the grid tariff. All actions satisfy physical feasibility (e.g., quantity bounded by forecast net position; SOC bounds for storage).

• Transition $T$. Deterministic: CDA clearing updates trades and inventories; the grid settles residual imbalances; storage decisions update SOC.

• $R$ is the fairness-shaped reward: raw economic profit or cost plus the LLM-derived bonuses, optionally adjusted for storage-related factors such as cycling cost or self-sufficiency incentives.

Details of the storage model, including SOC update, charge/discharge efficiencies, and power/capacity constraints are provided in Section 2.1, with scenario-specific penetration and parameter values in Section 3.

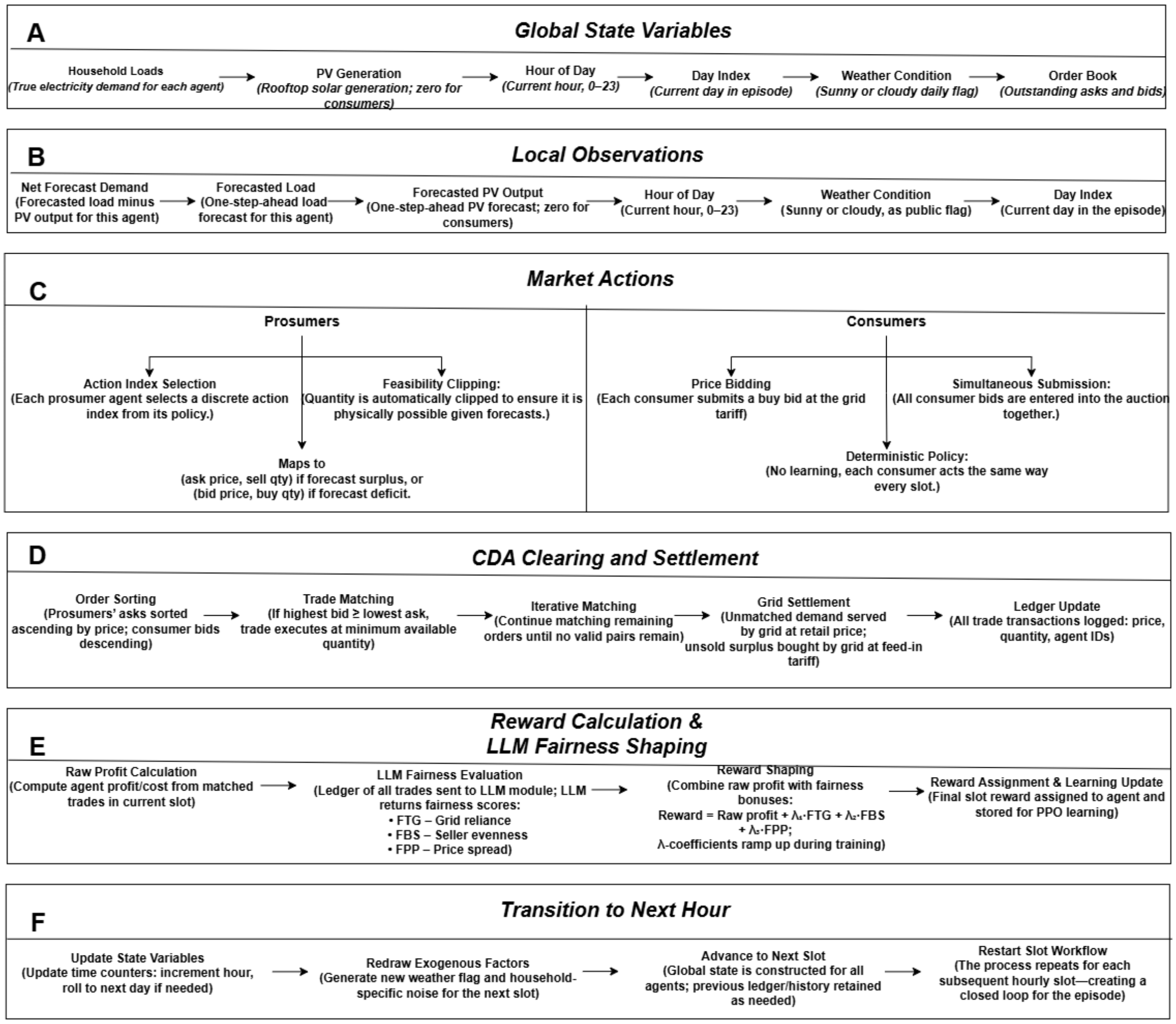

**Figure 2.** State–action–reward flow for a single hourly slot in the FairMarket-RL P2P market.

### 2.1 Environment Dynamics:

At the start of each hour $t$, the simulator generates each household's true electricity demand $L_{a,t}$ and, if applicable, rooftop-PV output $G_{a,t}$. Both are obtained by scaling canonical, unit-peak 24-hour templates with household-specific peaks, a daily weather intensity (via $\kappa_{d_t}$ from Eq. (1)), and small multiplicative noise to capture intra-day variability [31], [32]. We assume an hourly cadence $\Delta t = 1\, h$ and energy units in $kWh$:

$$L_{\{a,t\}} = \bar{L}_a \cdot \varphi_{load(h_t)} \cdot \eta^{L}_{a,t} \tag{4}$$

$$G_{a,t} = o_a \cdot \bar{G}_a \cdot \kappa_{d_t} \cdot \varphi_{pv(h_t)} \cdot \eta^{G}_{a,t} \tag{5}$$

- $\bar{L}_a$ and $\bar{G}_a$are the peak hourly demand and peak $PV$ yield for household $a$;
- $\varphi_{load(\cdot)}$ and $\varphi_{pv(\cdot)}$ are normalized 24-hour profiles (unit peak) evaluated at $h_t \in \{0, \dots, 23\}$;
- $\kappa_{d_t}$ is the daily weather intensity from $Eq.$ (1);
- $o_a \in \{0,1\}$ is the $PV$ -ownership indicator; and $\eta^{L}_{a,t}, \eta^{G}_{a,t}$ are independent and identically distributed log-normal noise terms (mean 1, s.d. 5 %), independent across agents and time.

#### 2.1.1 Storage state and dynamics

For agents with batteries ( $s_a = 1$ ), the state-of-charge $B_{a,t}$ (kWh) is updated by hourly charge/discharge decisions and projected to capacity bounds:

$$\tilde{B}_{a,t+1} = B_{a,t} + \eta_c \cdot q^{ch}_{a,t} - \left(\frac{q^{dis}_{a,t}}{\eta_d}\right) \tag{6}$$
$$B_{\{a,t+1\}} = \min\left\{ B^{max}_a, \max\left\{ 0,\ \tilde{B}_{a,t+1}\right\}\right\}$$

with power-rate limits ($for\ \Delta t = 1\, h$):

$$0 \le q^{ch}_{a,t} \le P^{ch,max}_a \cdot \Delta t, \tag{7}$$
$$0 \le q^{dis}_{a,t} \le P^{dis,max}_a \cdot \Delta t$$

where $q^{ch}_{a,t}$, $q^{dis}_{a,t}$ are energy (kWh), $\eta_c, \eta_d \in (0,1]$ are charge/discharge efficiencies, and $B^{max}_a$ is capacity (kWh) [37], [38]. If $s_a = 0$, storage variables are inoperative.

#### 2.1.2 Net quantity brought to market

The storage-adjusted net demand submitted for clearing is

$$D^{net}_{a,t} = L_{a,t} - G_{a,t} - s_a \cdot q^{dis}_{a,t} + s_a \cdot q^{ch}_{a,t} \tag{8}$$

This single expression covers consumers ($o_a = 0$) and non-storage agents ($s_a = 0$).

### 2.1.3 Forecasts and local observation

At the start of each slotuseholds receive one-step-ahead forecasts of their own load and (if applicable) PV output, modeled via unbiased Gaussian noise with 5% standard deviation [35], [36]:

$$\hat{L}_{a,t} = L_{a,t} \cdot \varepsilon^{L}_{a,t}, \quad (9)$$
$$\hat{G}_{a,t} = G_{a,t} \cdot \varepsilon^{G}_{a,t}, \quad \varepsilon^{(\cdot)}_{a,t} \sim N(1, 0.05^2).$$

When prosumer $i$ acts, the local observation includes its net forecast, component forecasts, time/weather, and (if storage-enabled) SOC:

$$s_t^{s_i} = \left[\hat{L}_{i,t} - \hat{G}_{i,t}, \quad \hat{L}_{i,t}, \quad \hat{G}_{i,t}, \quad s_i \cdot B_{i,t}, \quad h_t, \quad \kappa_{d_t}, \quad d_t\right] \quad (10)$$

The term $s_i \cdot B_{i,t}$ automatically zeroes the SOC component when no storage is present. This captures both forecast uncertainty and the inter-temporal coupling induced by storage [37], [38].

### 2.1.4 Action space and feasibility

Prosumers (two-sided participation): A prosumer may submit an ask (sell order) when it forecasts surplus and/or a bid (buy order) when it forecasts deficit, using discrete price–quantity menus that reflect retail-scale practice [39]. Prices are chosen from a fixed grid (default $10 - 30$ ¢ $kWh^{-1}$); quantities come from agent-specific menus bounded by $Q_i^{sell,max}$r asks and $Q_i^{buy,max}$r bids. Feasibility is enforced against the agent's own forecasts:

Ask feasibility:

$$0 \le q_{i,t}^{ask} \le \max\left(0, \quad \hat{G}_{i,t} - \hat{L}_{i,t}\right). \quad (11)$$

Bid feasibility:

$$0 \le q_{i,t}^{bid} \le \max\left(0, \quad \hat{L}_{i,t} - \hat{G}_{i,t}\right). \quad (12)$$

Consumers: Submit bids at the grid tariff, with quantities bounded by their forecast demand.

### 2.1.5 CDA clearing and transition

After asks and bids are submitted, a CDA greedily matches the highest bids (from consumers and prosumers) to the lowest asks (from prosumers) while the trade condition holds $\left(p_{buy} \ge p_{sell}\right)$; matched quantities execute at the minimum available volume, orders are resized/removed, and the loop continues until no executable pair remains [41]. The grid settles residual imbalances at its retail/feed-in tariffs. The transition T is deterministic given the actions: CDA updates trades/inventories, storage actions update SOC via Eq. (6), and time/weather advance to the next slot.

### 2.2 Raw Reward Formulation

Once trades are cleared up for the current hour, each agent receives an immediate economic reward. For prosumers the reward is the difference between sales revenue and purchase expenditure, while for consumers it is simply the cost of energy acquired. These are the raw rewards before any fairness shaping is applied.

For a prosumer $(seller)\ i \leq m$, let $(i \rightarrow \kappa)$ denote every trade where $i$ sells energy to some counterparty, and $(\kappa \rightarrow i)$ every trade where $i$ buys energy. The hourly profit is:

$$\pi_{i,t} = \Sigma_{\{(i \rightarrow \kappa)\}p} \cdot q \ - \ \Sigma_{\{(\kappa \rightarrow i)\}p} \cdot q \qquad (13)$$

Over an entire episode of $H = 24D$ slots, the raw return to prosumer $i$ is:

$$r_i^{raw} = \Sigma_{\{t=0\}_{\{i,t\}}^{\{H-1\}\pi}} \qquad (14)$$

For a consumer $j > m$, every trade $(\kappa \rightarrow j)$ represents energy purchased from counter-party $\kappa$. The hourly cost is:

$$\chi_{j,t} = \Sigma_{\{(\kappa \rightarrow j)\}p} \cdot q \qquad (15)$$

The corresponding raw return (negative, because it is a cost) across the episode is:

$$r_j^{\{raw\}} = -\Sigma_{\{t=0\}_{\{j,t\}}^{\{H-1\}\chi}} \qquad (16)$$

These raw returns serve as the baseline economic signals before we incorporate the LLM-based fairness bonuses described in the next section. Note that because each hour is independent (no storage), the sums decouple across time.

### 2.3 LLM-Based Fairness Shaping

Raw economic rewards capture profitability but may encourage behavior that is socially sub-suboptimal, for example, relying too heavily on the grid or allowing a single prosumer to dominate trade volume. To realign incentives, we send the post-slot ledger to LLM, which returns three fairness metrics scaled to the interval [0, 1]:

- FTG– Grid reliance (1 if all demand met via P2P, 0 if entirely grid-supplied).
- FBS – Fairness between sellers (1 when quantities are evenly shared).
- FPP – Fair pricing spread (1 when clearing prices cluster tightly around the median).

For each prosumer $i$ the slot-level reward is:

$$R_{i,t} = \pi_{i,t} + \lambda_{\{grid\}(e)} \cdot \beta_{grid} \cdot FTG_t + \lambda_{\{price\}(e)} \cdot \beta_{price} \cdot FPP_t + \lambda_{\{peer\}(e)} \cdot \beta_{peer} \cdot FBS_t \cdot \left( \frac{q_{i,t}^{sold}}{\sum_{k=1}^{m} q_{k,t}^{sold}} \right) \quad (17)$$

The first term is the prosumer's raw profit $\pi_{\{i,t\}}$. The next two terms reward low grid reliance and tight price spreads, while the final term encourages equitable sharing among sellers the factor $\frac{q_{i,t}^{sold}}{\sum_{k=1}^{m} q_{k,t}^{sold}}$ scales the bonus by the seller's contribution in that slot and $\beta_{grid}, \beta_{price}, and\ \beta_{peer}$ are scaling factors.

Each fairness-weight coefficient $\lambda_{\bullet(e)}$ is ramped linearly during training to prevent fairness bonuses from overpowering raw economic incentives in the early episodes. The schedule is defined piece-wise:

(a) Inactive segment:

$$\lambda_{\bullet(e)} = 0 \qquad for \quad e < e_{\bullet}^{start} \quad (18)$$

(b) Linear ramp:

$$\lambda_{\bullet(e)} = \frac{(e - e_{\bullet}^{start})}{(e_{\bullet}^{full} - e_{\bullet}^{start})} for \quad e_{\bullet}^{start} \leq e < e_{\bullet}^{full} \quad (19)$$

(c) Full strength:

$$\lambda_{\bullet(e)} = 1 \qquad for \quad e \geq e_{\bullet}^{full} \quad (20)$$

We denote each fairness-weight coefficient as $\lambda_{\bullet(e)}$, where e is the training episode index and $\bullet \in \{grid, spread, peer\}$ indicates the fairness metric type. Training spans $E = 10{,}000$ episodes. The grid-reliance and price-spread weights ramp linearly from $e = 0.02E$ to $e = 0.30E$, while the peer-fairness weight ramps from $e = 0.30E$ to $e = 0.80E$, allowing agents to learn basic trading before strong fairness incentives dominate.

## 3. Case Study

In the previous sections, we detailed the market rules, state transitions, reward mechanisms, and training regimen that form the core of the FairMarket-RL simulator. To progress from theoretical design to practical validation, we subject the trained FairMarket-RL agents to a series of controlled case studies. Each scenario fixes a specific combination of prosumer and consumer types, simulation horizon, and weather profile, then runs the market without further tuning of learning hyperparameters. Scalability is evidenced by the smooth transition from a 30-day, three-prosumer pilot to a 90-day, ten-prosumer community, confirming that the platform can expand without fundamental redesign.

### 3.1 Case 1- 30-day simulation community

#### 3.1.1 Experimental setup and inputs

We evaluate FairMarket-RL in a 30-day pilot community comprising three rooftop-PV prosumers ($P_1$–$P_3$), two consumers ($C_1$–$C_2$), and a single price-taking grid agent. The market operates as 720 hourly auctions cleared by a CDA. In each auction, participants submit discrete price–quantity menus, mirroring retail practice while preserving strategic flexibility. Policies are trained under hard constraints (price bounds, feasibility checks, gradient clipping) and soft targets (maintaining a target peer-to-grid balance, meeting fairness indices, capping individual seller shares).

Daily operating conditions are driven by solar variability and load. Figure 3 summarizes the inputs used in this case: average community load, PV generation, and net demand profiles for sunny versus cloudy days. On sunny days, midday PV averages approximately 12 kW and supplies about 51% of community load, often pushing net demand negative for multiple hours. On cloudy days, the midday PV peak averages approximately 4.4 kW and covers about 38% of load. In both regimes, evening peaks occur after sunset and must be met by imports (from peers or the grid).

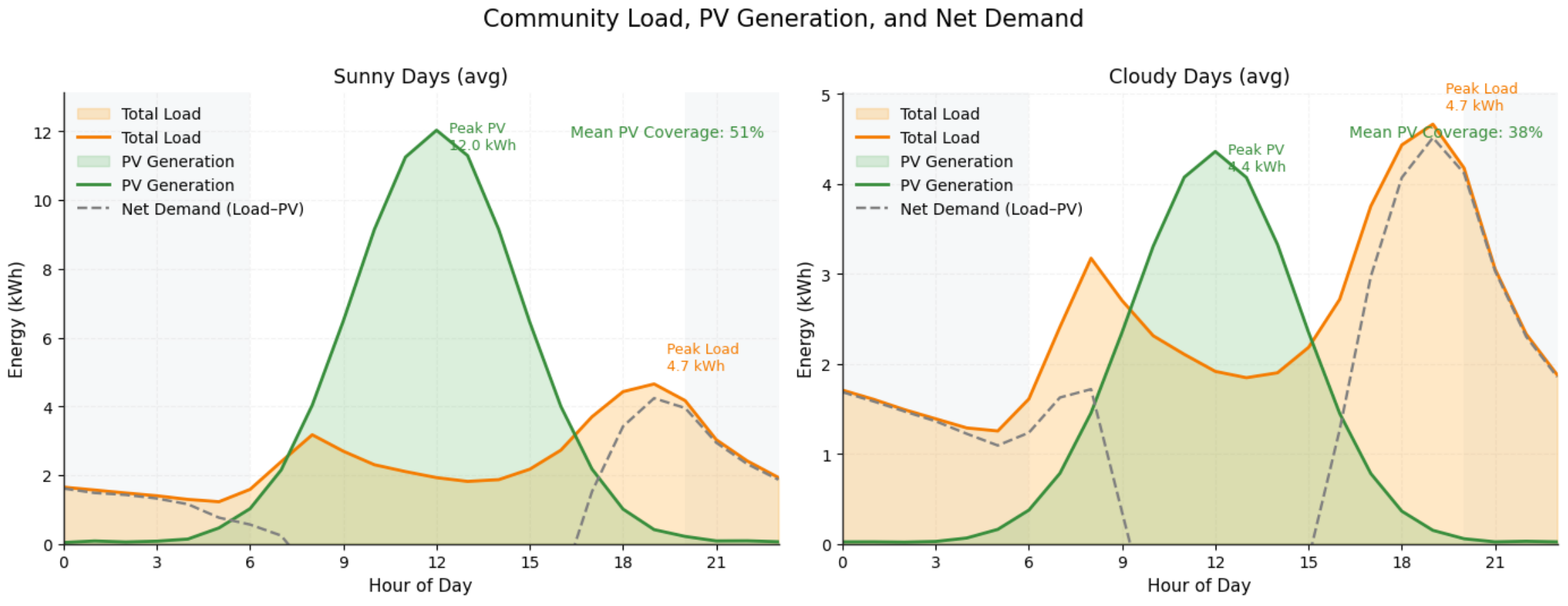

**Figure 3.** Average community load, PV generation, and net demand for sunny (left) and cloudy (right) days (30-day averages).

#### 3.1.2 Policy training: convergence and fairness

Over 10,000 PPO episodes, total rewards for $P_1$–$P_3$ rise sharply within approximately 1,000–1,500 episodes and then stabilize, indicating fast convergence. Fairness metrics remain high throughout training: FTG stabilizes around 0.80–0.85, FBS near 0.90, and FPP consistently exceeds 0.95. These trajectories show that the learned policies balance efficiency (high rewards) with equity (stable, high fairness).

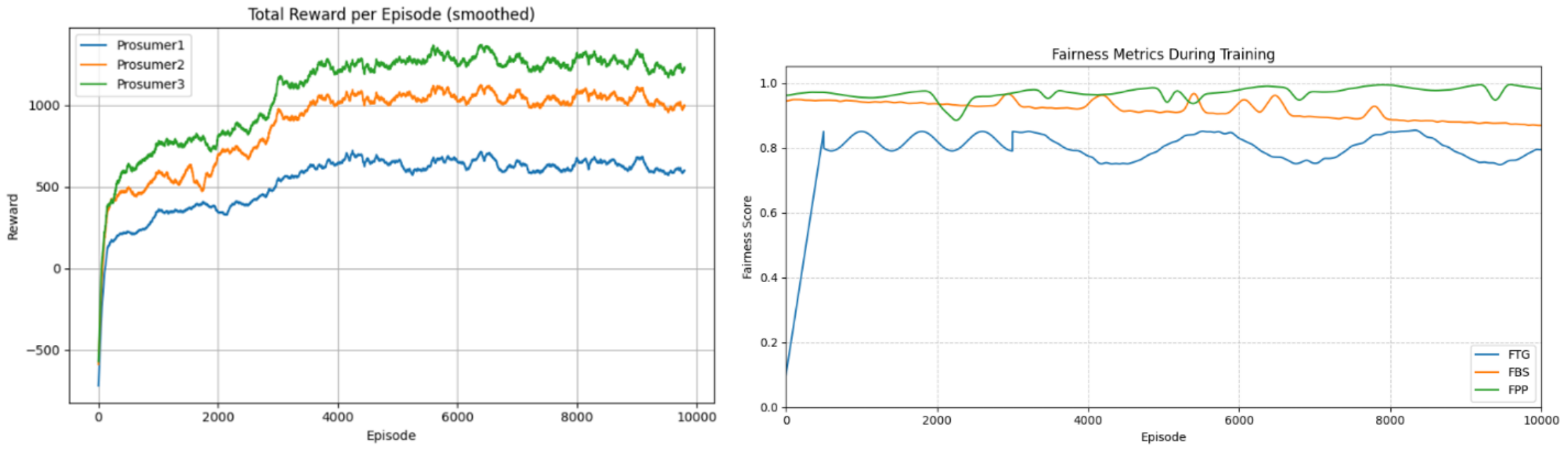

**Figure 4.** Case 1 — Training diagnostics. (Left) Total reward per episode for $P_1$–$P_3$; (Right) fairness metrics (FPP, FBS, FTG) over 10,000 training episodes.

### 3.1.3 Aggregate energy flows and diurnal operation

Across the 30-day horizon, the community transacted 1,026 kWh peer-to-peer and 874 kWh via the grid, yielding an approximately 54.0% / 46.0% peer-to-grid split. The largest single-hour peer trade reached about 24 kWh in the sunniest midday interval (an average power rate of approximately 24 kW). Instantaneous net power ranged from +4 kW surplus to –5 kW deficit and remained within operational limits. The complete 720-hour simulation executed in under 40 minutes on a standard 12-core CPU, indicating that continuous, day-level resolution is computationally tractable.

Trading exhibits a clear diurnal rhythm (Figure 5): daylight hours are dominated by peer-to-peer exchanges aligned with PV availability and tapering to near zero overnight. Grid imports mirror this pattern, rising after sunset and in the early morning. Short midday intervals with community-wide PV surplus yield modest exports to the grid.

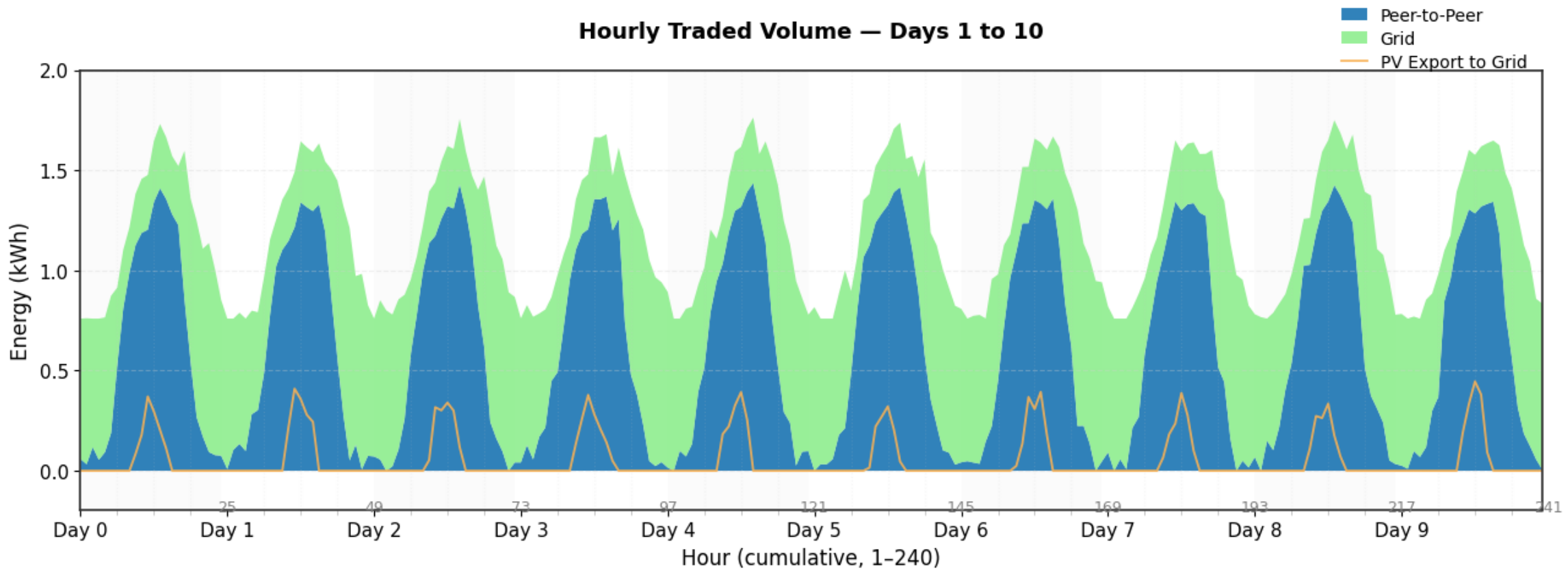

**Figure 5.** Case 1 f Hourly energy volumes (Days 1–10). Peer-to-peer exchanges, grid imports, and PV exports to the grid.

### 3.1.4 24-hour role-switching snapshot

A representative day (Figure 6) shows enforced hour by hour role exclusivity and the calibrated peer to grid split. From midnight through sunrise (about 07:00), the grid steadily supplies about 2 to 3.5 kWh per hour to cover residual demand and charging while prosumers mostly buy. As PV ramps up after sunrise, P1 to P3 switch into seller roles: P2 reaches a midday sell peak of approximately 9.95 kWh with another surge near 16:00, P1 peaks at approximately 6.81 kWh around 14:00, and P3 at approximately 5.93 kWh near 09:00 to 10:00. During these PV rich hours, consumers buy primarily from peers; C1 and C2 peak at approximately 9.01 kWh at 16:00 and approximately 5.70 kWh at 14:00, respectively. The grid also acts as a buyer (about 8 kWh total) by absorbing FIT PV exports. After sunset (about 17:00), prosumers revert to net buying and the grid again covers most demand.

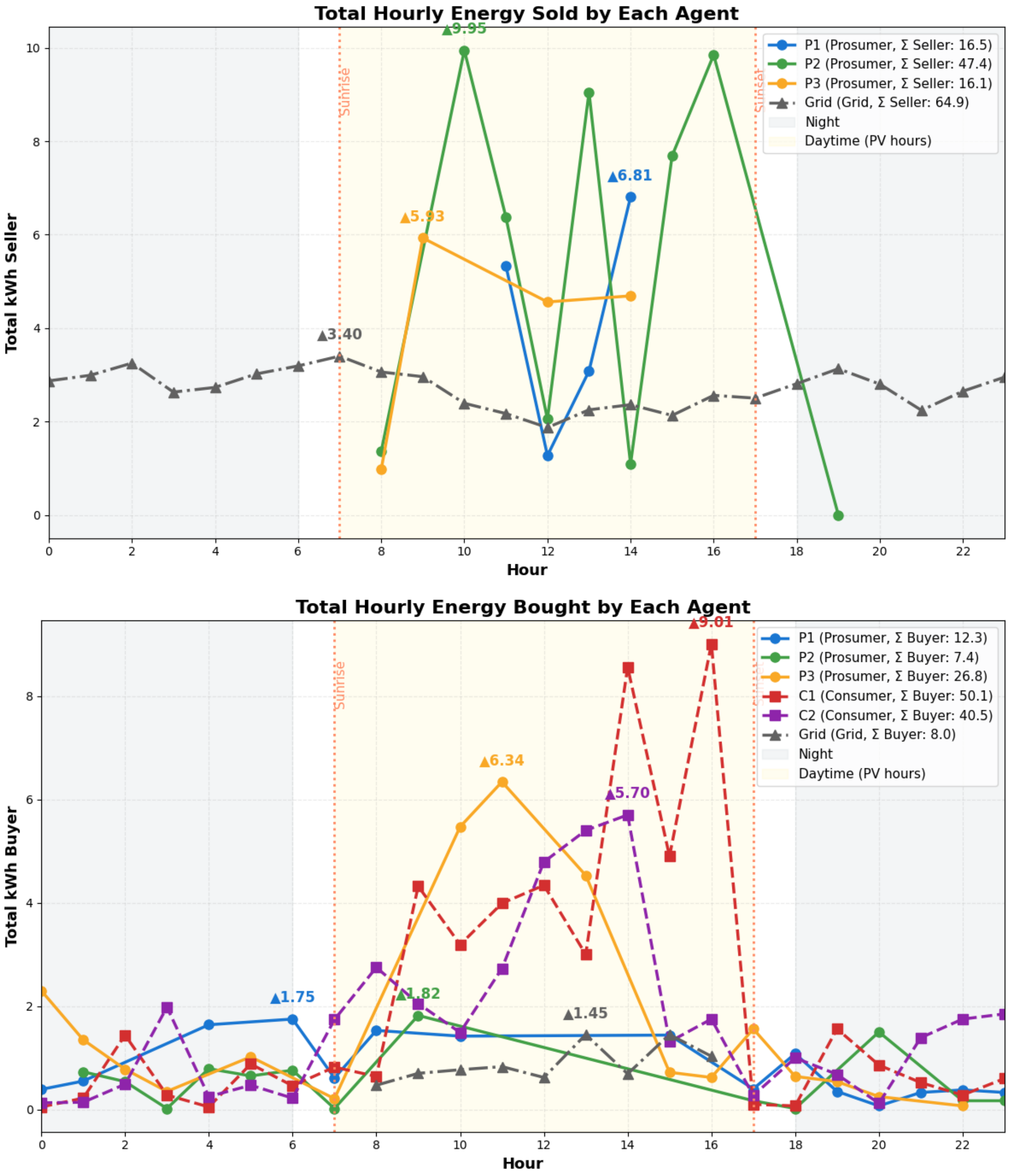

**Figure 6.** (Top) Total hourly energy bought by each agent (representative day). (Bottom) Total hourly energy sold by each agent (same day).

### 3.1.5 Community Economics

Over 30 days, $P_2$ attains the highest net profit ($176.9), followed by $P_1$ ($174.1) and $P_3$ ($169.5), driven primarily by peer sales with smaller grid exports. Consumers incur nearly identical total costs ($C_1$ $197.1, $C_2$ $196.3), with more than 60% of expenditure from peer purchases, evidencing access to lower-priced local energy. The grid remains financially viable with $238.2 in revenue, $180.5 in costs, and $57.7 net profit. Overall, the market sustains positive outcomes for all roles while preserving the approximately 54/46 peer-to-grid balance.

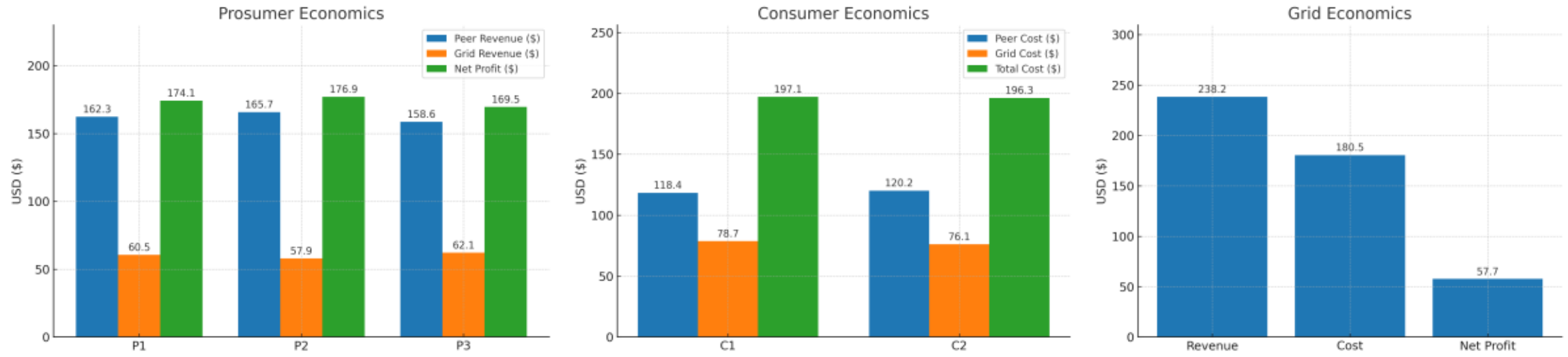

**Figure 7.** Case 1 — Economics (30 days). (Left) Prosumer net profit; (Center) Consumer total cost; (Right) Grid revenue, cost, and net profit.

## 3.2 Case 2 — 90-Day Scaled Community

### 3.2.1 Experimental setup and inputs

We extend FairMarket-RL from the Case 1 pilot to a larger mixed-asset community comprising ten rooftop-PV prosumers ($P_1$–$P_{10}$), three consumers ($C_1$–$C_3$), and a single price-taking grid node. The market operates as 2,160 hourly auctions over 90 days, cleared by the same CDA. In each auction, participants submit discrete price–quantity menus, mirroring retail practice while preserving strategic flexibility. Physical constraints (power balance, feeder limits) and soft targets (peer-to-grid energy ratio, fairness indices, seller-share caps) are identical to Case 1. Daily operating conditions follow a seasonally balanced solar-irradiance profile with community load. Fairness-shaped PPO policies trained in Case 1 are transferred without retraining, allowing the study to isolate the effects of greater market liquidity, more diverse generation–demand patterns, and a longer operating horizon on performance and fairness.

### 3.2.2 Policy validity: convergence and fairness

To document policy behavior in the scaled environment, we report Case-2 diagnostics in Figure 8. Episode returns exhibit rapid early gains and stabilization, consistent with Case 1. Fairness remains high across episodes: FPP stays above 0.95, FBS remains near 0.90, and FTG improves to about 0.72 by late training. These plots confirm that the transferred policies preserve efficiency and equity under increased market liquidity and a longer horizon.

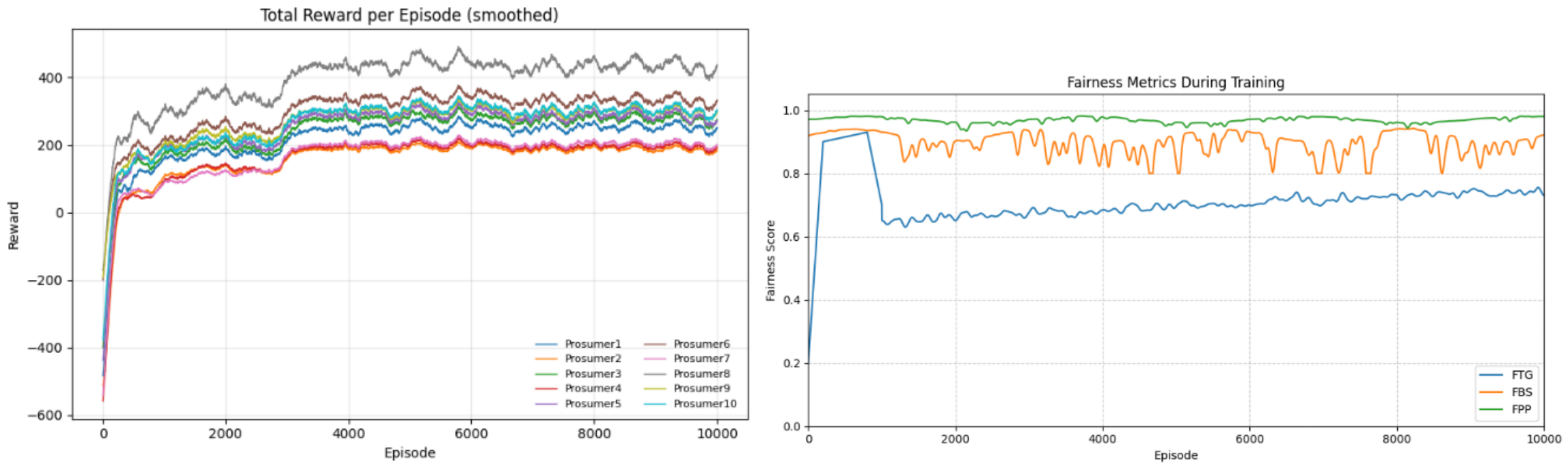

**Figure 8.** Case 2 — Training diagnostics. (Left) Total reward per episode for $P_1$–$P_{10}$; (Right) fairness metrics (FPP, FBS, FTG) over 10,000 training episodes.

### 3.2.3 Aggregate energy flows, diurnal patterns, and scalability

Over 90 days, the community transacted 2,260 kWh peer-to-peer and 1,960 kWh via the grid, yielding an approximately 53.6% / 46.4% peer-to-grid split. The largest single-hour peer trade reached about 2.4 kWh in the sunniest midday interval (an average power rate of approximately 2.4 kW for that hour). Instantaneous net power ranged from +6 kW surplus to –8 kW deficit and remained within operational limits. The 2,160-hour simulation completed in about 2 hours 15 minutes on a standard 12-core CPU, demonstrating practical scalability for continuous, quarter-season horizons. Peer-to-peer volumes track solar availability, peaking at about 2.4 kWh near midday and declining after sunset (Figure 9). Grid imports dominate night-time hours and two brief cloudy intervals. Over Days 1–10, mean hourly volumes are about 1.05 kWh (peers) and about 0.91 kWh (grid), consistent with the long-run 90-day split.

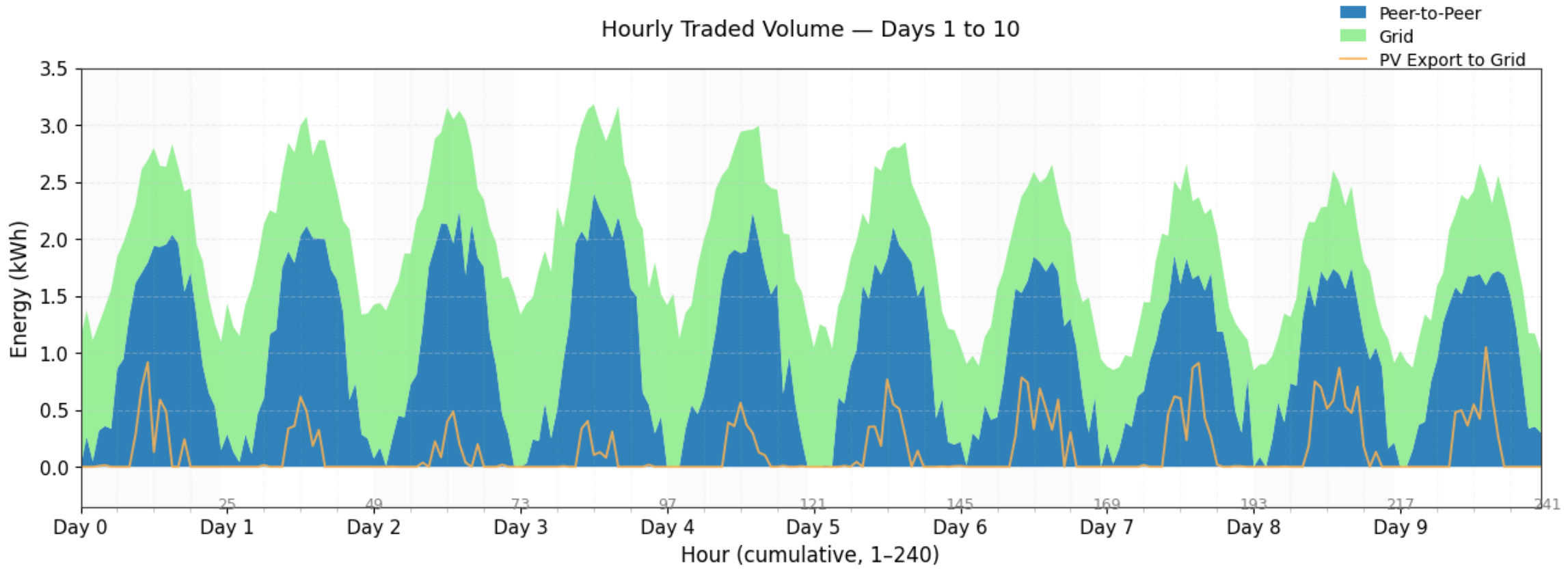

**Figure 9.** Case 2 — Hourly energy volumes (Days 1–10). Peer-to-peer exchanges, grid imports, and PV exports to the grid.

### 3.2.4 Market quality and inclusivity

Market health is assessed using three indicators (Figure 9). Clearing-price spreads range between 10 and 22 ¢/kWh, with a smoothed median near 18 ¢/kWh wider during cloudy, supply-constrained periods and narrower under clear-sky, high-liquidity conditions. Seller-share entropy lies between 0.60 and 0.96 (mean about 0.78), indicating that no single prosumer dominates sales. Jain's fairness index (JFI) stays above 0.90 for roughly two-thirds of hours and never falls below 0.70, confirming equitable allocation despite variability in generation and demand.

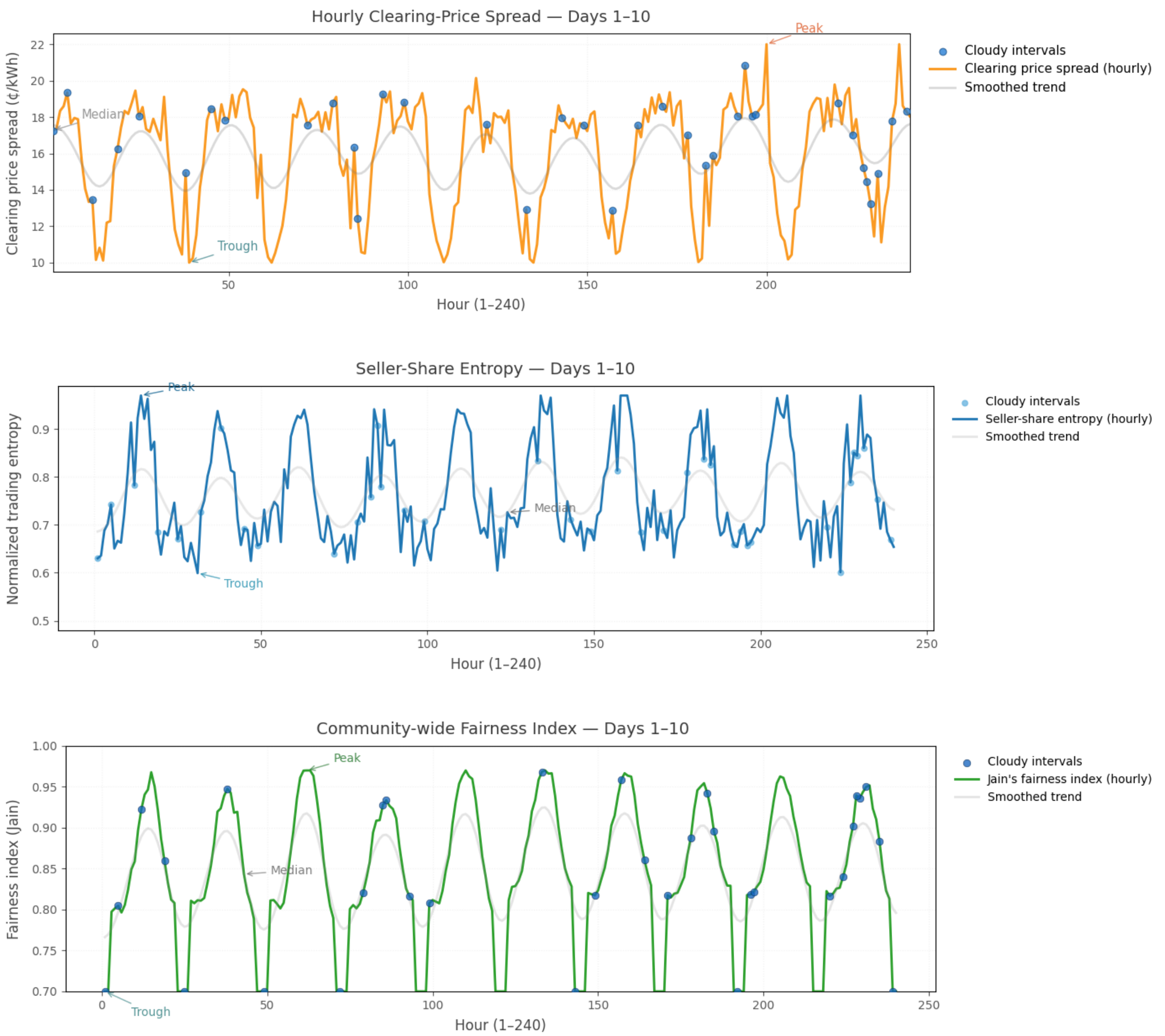

**Figure 9.** Hourly market fairness metrics for Case 2 over Days 1–10: (top) clearing-price spread, (middle) seller-share entropy, (bottom) JFI.

### 3.2.5 24-hour role-switching snapshot

A representative day (Figure 10) shows the enforced hour by hour exclusivity. Purchasing (top): the grid is the dominant buyer overnight and near sunrise, with an hourly peak of approximately 1.8 kWh, then its share declines as peer trades take precedence. Consumers C1 to C3 keep steady demand across the day, totaling about 6.6, 6.7, and 7.5 kWh, respectively. Among prosumers, P10 is the largest buyer at about 3.9 kWh, followed by P4 at about 3.6 kWh, with P3 and P7 at about 3.1 kWh each. Selling (bottom): during daylight, prosumers switch into seller roles. P2 is the leading supplier at about 6.8 kWh for the day; the next largest are P5 at about 4.2 kWh, P6 at about 4.1 kWh, and P1 at about 4.0 kWh, with others contributing between about 1.4 and 3.5 kWh. The grid sells approximately 21.8 kWh, mainly covering morning and evening deficits. The clear role reversals around sunrise and sunset illustrate adaptive balancing between self-consumption

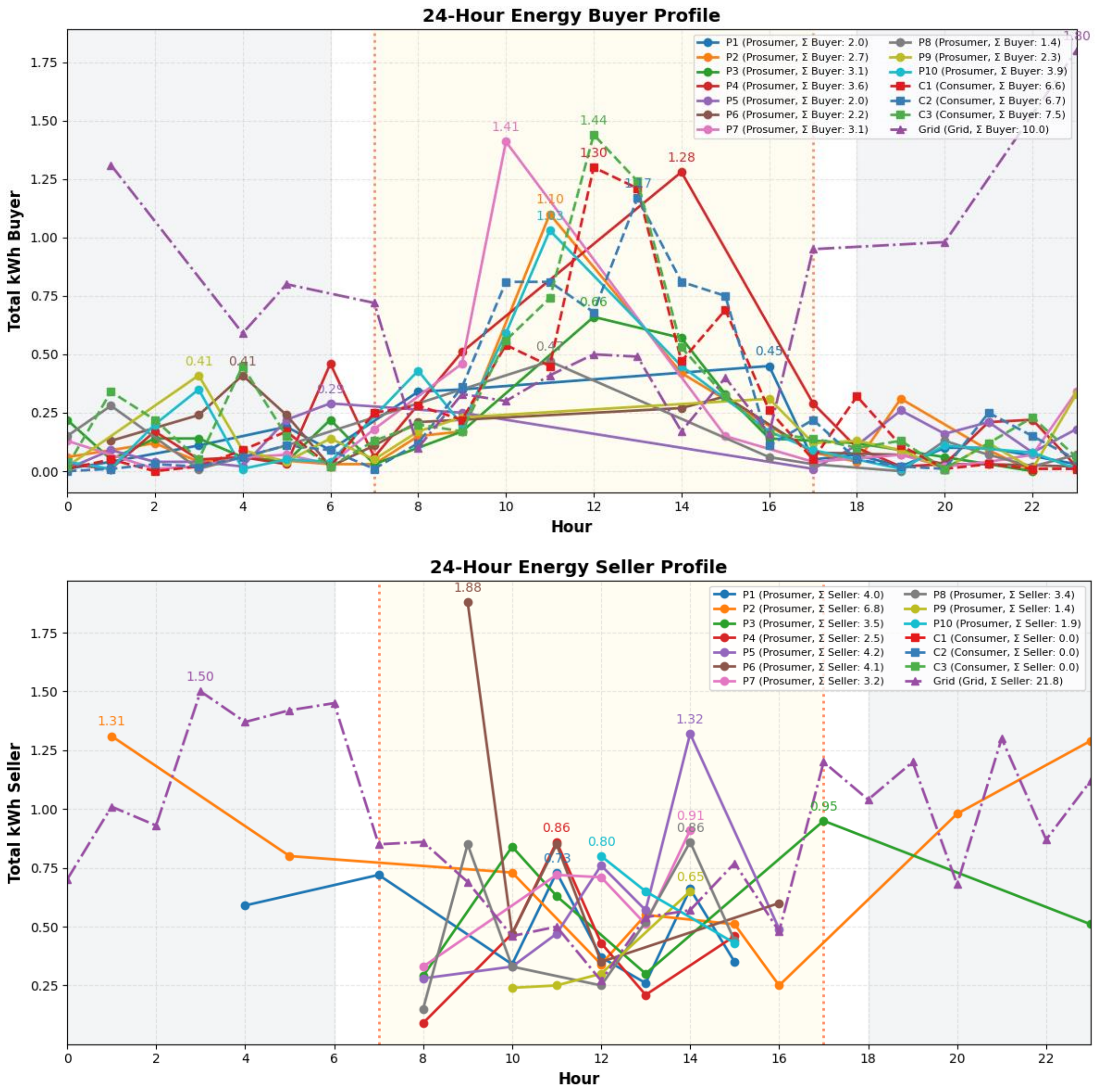

**Figure 10.** Hourly energy purchased (top) and sold (bottom) by each agent on a representative day.

### 3.2.6 Community Economics

Consumers realize substantial bill reductions with average monthly costs of \$295 ($C_1$), \$310 ($C_2$), and \$325 ($C_3$) about 27–28% lower than grid-only baselines. Prosumers earn an average net profit of about \$160 (range \$120–\$220) primarily from peer sales and modest feed-in tariffs. The grid remains profitable with about \$710 in revenue and about \$120 in operating costs, for about \$590 net profit over the 90 days. Collectively, the system maintains the approximately 54/46 peer-to-grid balance while delivering positive economic outcomes for all participant classes.

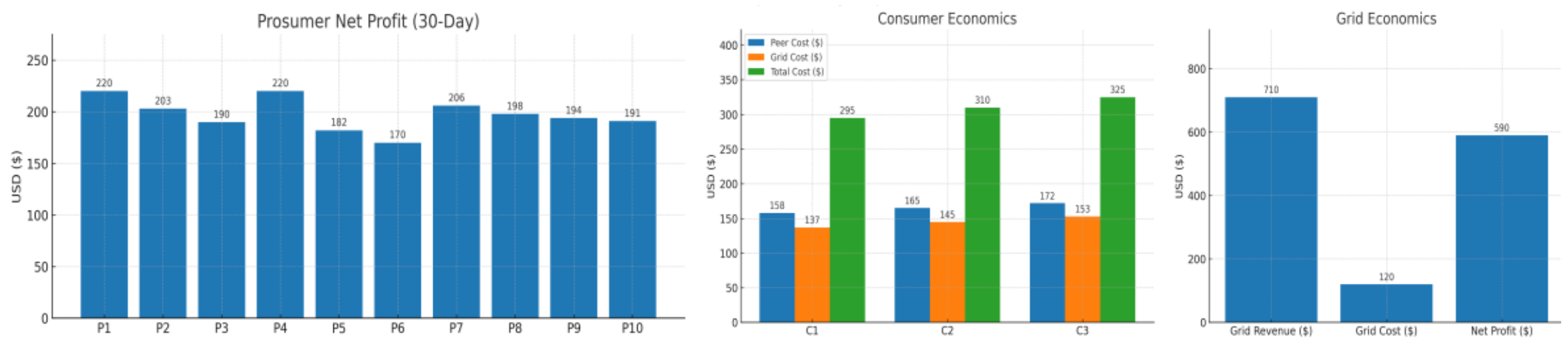

**Figure 11.** Case 2 — Economics (90 days). (Left) Prosumer net profit; (Center) Consumer average monthly bill and source composition; (Right) Grid revenue, cost, and net profit.

### 3.3 Case 3- 90-day real world community

### 3.3.1 Experimental setup and inputs

This final case evaluates FairMarket-RL under realistic operating conditions using a complete, real-world community dataset. We use "A complete energy-community dataset with photovoltaic generation, battery energy-storage systems, and electric vehicles (v1.5)" collected from residential households across Europe. Twelve representative households are selected to reflect typical diversity: three battery-equipped PV prosumers ($P_1$–$P_3$), seven PV-only prosumers ($P_4$–$P_{10}$), and three pure consumers ($C_1$–$C_3$). The original 15-minute load and PV series are aggregated to hourly values to align with market operations. The market runs as hourly continuous double auctions (CDA) over 90 consecutive days. In each auction, agents submit discrete price–quantity menus. We include dynamic grid prices (imports/exports), battery state-of-charge limits, and price-feasibility bounds. All agents use pre-trained PPO policies with fairness shaping. Policies are fixed from previous cases checkpoints to evaluate generalization without retraining or hyperparameter changes.

### 3.3.2 Convergence and fairness

The training diagnostics for Case 3 present the evolution of total episode rewards and fairness metrics. In the training plots, the left panel shows that all three battery-equipped prosumers rapidly increase their cumulative rewards during the early episodes, with each agent's performance plateauing as learning stabilizes. Notably, Prosumer 3 achieves the highest average reward, followed by Prosumer 2 and then Prosumer 1, suggesting that differences in load patterns or PV production created slightly different opportunities for profit and arbitrage. The seven PV-only prosumers display the same qualitative behavior monotonic reward growth and stabilization within the same episode window with lower variance owing to the simpler action space. The learning curves indicate that all agents learned to exploit the market structure and battery flexibility to maximize returns, with diminishing volatility as policy convergence was achieved.

The right panel tracks the evolution of the three primary fairness metrics over the course of training. FPP remains near its target level, indicating that market participation is balanced across agents. Both FBS and FTG improved substantially in the first 1,000 episodes, with FBS stabilizing well above the 0.8 fairness threshold and FTG converging slightly below it. These results demonstrate that reward shaping guides agents toward behaviors that not only maximize profit but also promote equitable access to market participation and benefits. In summary, the training process aligns economic incentives with the desired fairness properties, as evidenced by robust learning dynamics and stable, high fairness scores.

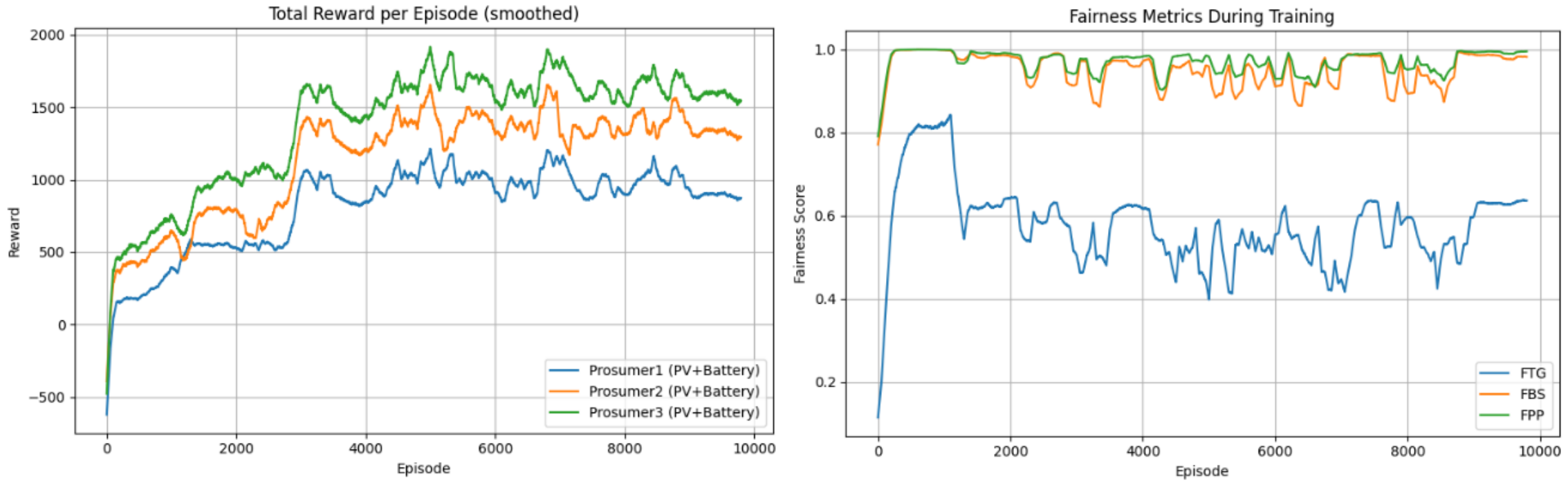

**Figure 12.** Case 3 — Training diagnostics. (Left) Total reward per episode for battery-equipped prosumers; (Right) fairness metrics (FPP, FBS, FTG) over 10,000 training episodes.

### 3.3.3 Sensitivity analysis (robustness to PV and load shocks)

We evaluate resilience via four counterfactuals with fixed policies: ±20% PV and ±10% aggregate load. A 20% PV increase expands peer trading to 5,782 kWh (+18%), contracts grid imports by 9%, raises aggregate prosumer profit by 11%, and reduces consumer bills by 6%; trading entropy rises to about 0.92, suggesting surplus is widely shared. A 20% PV shortfall reduces peer volume by 17%, increases grid reliance by 11%, raises consumer expenditure by 7%, and lowers prosumer profit by 10%; entropy remains about 0.81, showing equitable access despite scarcity. Higher demand (+10% load) drives an 11% increase in grid trades and a 9% rise in consumer costs, while peer activity and prosumer earnings tick up (+5% and +2%); entropy softens slightly (about 0.85). Lower demand (–10% load) trims grid imports by 8% and consumer costs by 7%, modestly reducing grid profit (about –5%) while leaving peer equity virtually intact (entropy about 0.90).

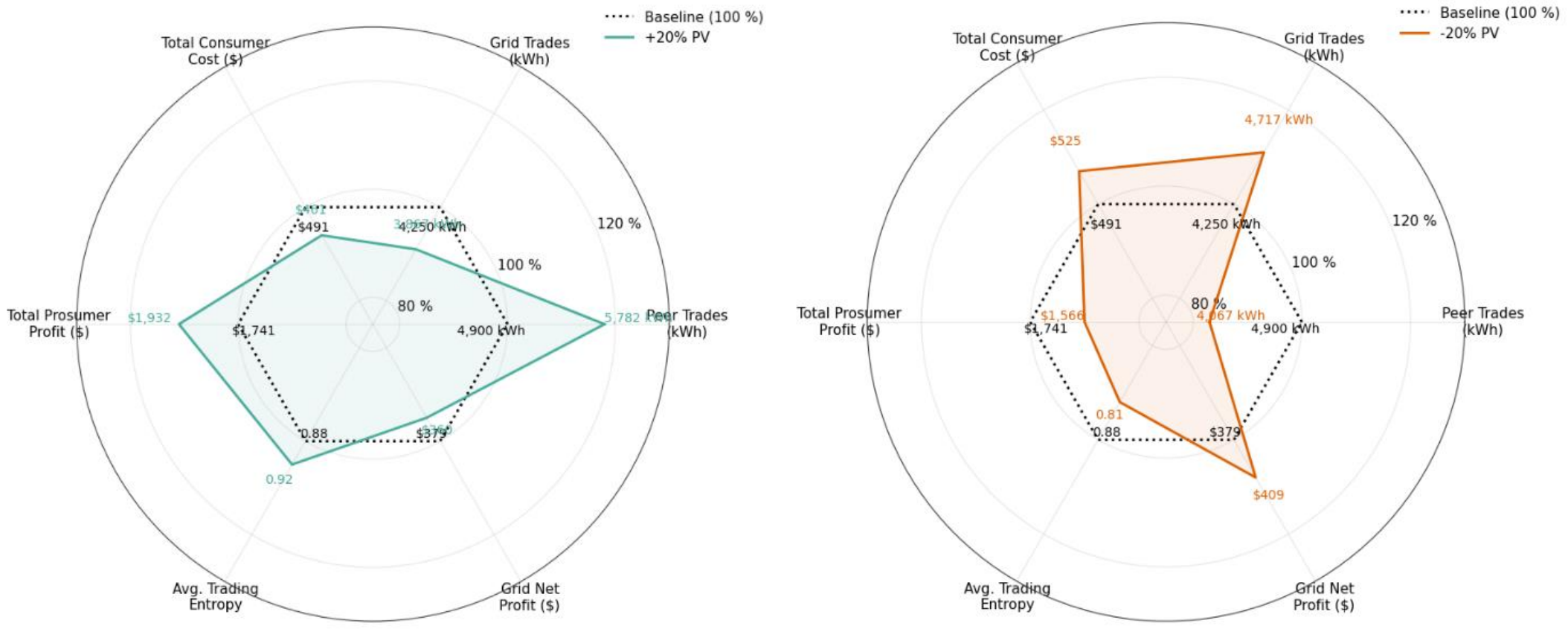

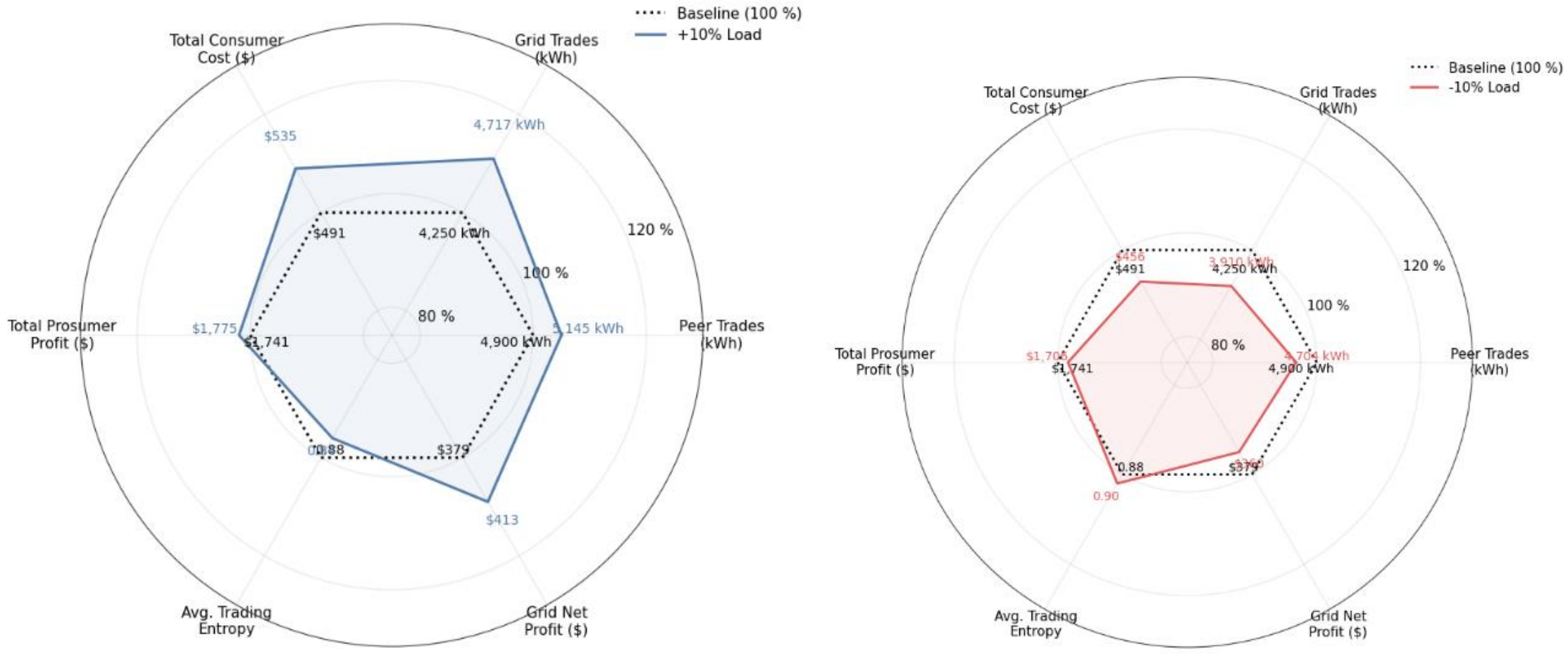

**Figure 13.** Sensitivity-analysis radar plots: impact of ±20% PV (panels a, b) and ±10% load (panels c, d) on peer trades, grid trades, consumer cost, aggregate prosumer profit, average trading entropy, and grid net profit.

### 3.3.4 Operational dynamics with batteries

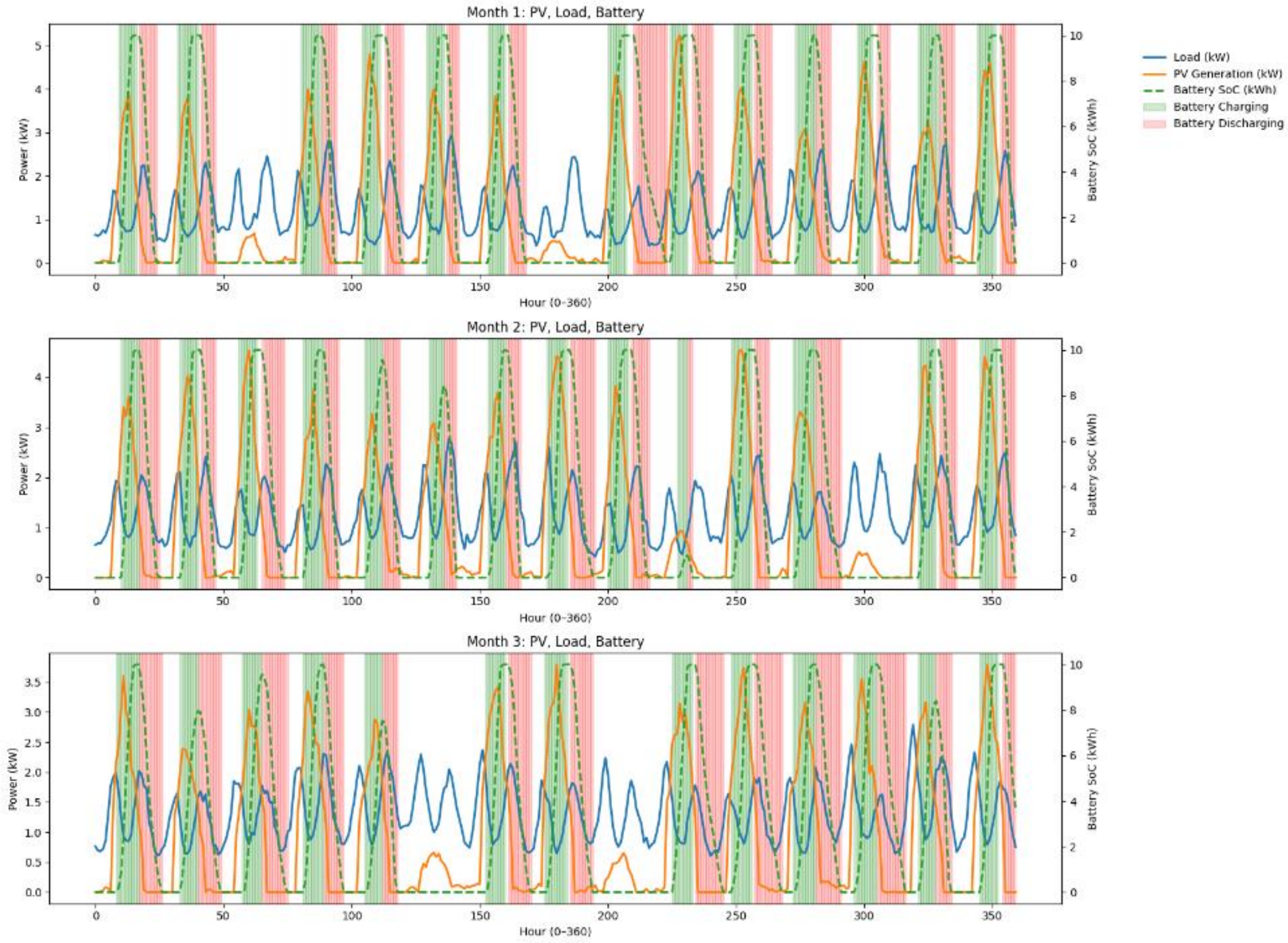

**Figure 14.** Daily battery charge–discharge cycles and state of charge across three representative weeks (Months 1–3).

Time-series profiles for a representative battery-equipped prosumer over three 15-day windows (distinct months) show seasonal and weather-driven variability. PV peaks decline from approximately 5 kW (Month 1) to below 4 kW (Month 3); load exhibits morning/evening peaks that drive cycling. In Month 1, daytime surpluses regularly saturate a 10-kWh battery with brief evening discharges; Months 2–3 show fewer charges and longer discharges approaching empty, especially during storms. This progression marks a shift from frequent self-sufficiency toward greater grid reliance later in the season.

### 3.3.5 Energy routing under market and storage design

We contrast three configurations over 15 days: (i) no P2P trading, (ii) P2P without batteries, and (iii) P2P with batteries. Without P2P, evening deficits are largely met by grid imports even when daytime PV is abundant. Enabling P2P redistributes daytime PV surpluses locally and reduces grid dependence. Adding batteries extends peer-supplied energy into evening and night, flattening grid imports to residual needs; most PV surplus is self-consumed or traded locally.

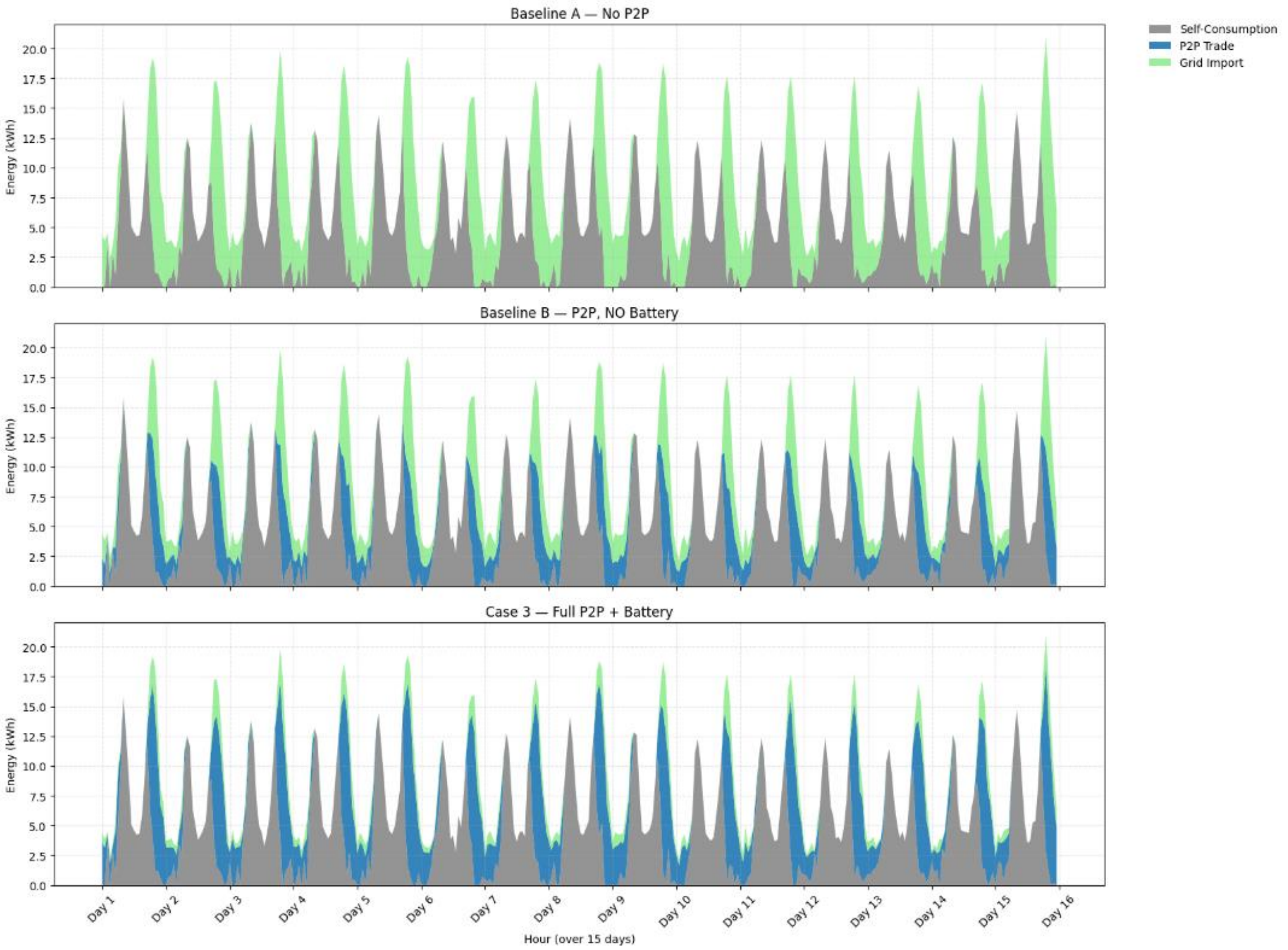

**Figure 15.** Scenario comparison – Community energy flows (15 days). (Top) No P2P; (Middle) P2P without batteries; (Bottom) P2P with batteries.

### 3.3.6 Community Economics

Battery-equipped prosumers achieve the largest net profits via arbitrage and self-consumption; PV-only prosumers maintain positive margins through local sales; consumers see sustained bill reductions (often exceeding 30% versus grid-only). The grid operator retains a positive net profit (about $379), indicating compatibility between decentralized peer trading and utility viability.

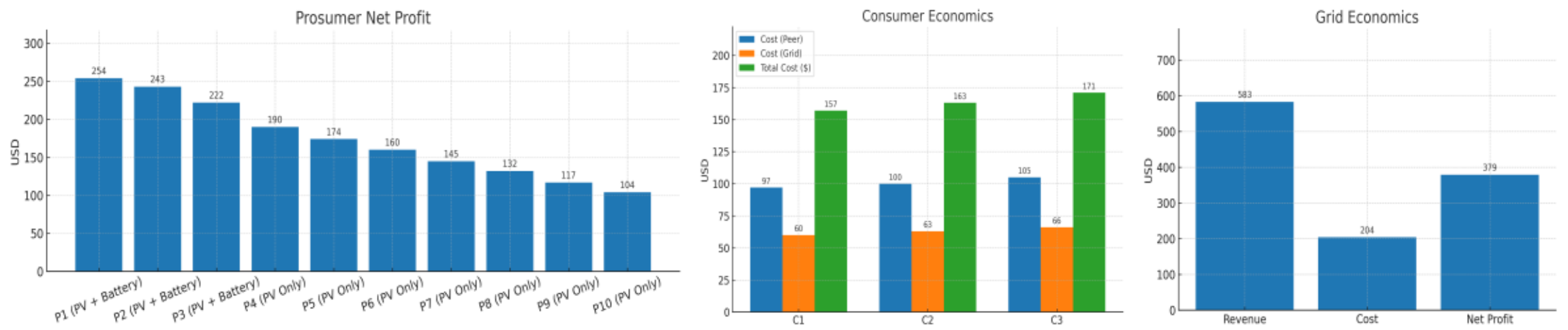

**Figure 16.** Case 3 — Economics (90 days). (Left) Prosumer net profit; (Center) Consumer average monthly bill and source composition; (Right) Grid revenue, cost, and net profit.

Across 90 days, peers exchanged approximately 4,900 kWh while grid transactions totaled about 4,250 kWh, a 54% / 46% peer-to-grid split reflecting a local-first preference without compromising reliability. Battery-equipped prosumers ($P_1$–$P_3$) achieved around $240 per agent; PV-only prosumers ($P_4$–$P_{10}$) earned about $160 per agent; consumers ($C_1$–$C_3$) reduced bills by about 30% versus grid-only baselines. Average trading entropy of about 0.88 confirms broad participation despite asset heterogeneity, and the grid remained economically viable with approximately $379 net profit. Robustness tests show proportional responses to PV and load shocks while equity and grid viability remain within targets. Computation is practical (full 90-day run in about three hours on a standard 12-core workstation), supporting planning studies and operational what-if analyses.

## 4.Conclusion

FairMarket-RL demonstrates that LLM-guided fairness shaping can be embedded into multi-agent reinforcement learning for peer-to-peer electricity markets without sacrificing economic performance or training stability. By coupling discrete price–quantity bidding in a continuous double auction with post-slot fairness feedback (FTG, FBS, FPP), the framework aligns private incentives with community-level equity under partial observability. The design is modular and physically grounded enforcing price and feasibility limits, stabilizing learning via PPO gradient clipping, and, when storage is present, handling inter-hour state-of-charge dynamics within the same action–observation scaffold. Evaluations spanning small pilots, larger synthetic communities, and a real-world mixed-asset dataset show consistent qualitative gains: trades shift toward local exchanges, consumer costs fall relative to grid-only procurement, seller participation remains balanced, and utility viability is preserved. Sensitivity analyses further indicate that fairness shaping acts as a stabilizing influence, with policies adapting smoothly to variation in solar availability and aggregate demand. Collectively, these results position FairMarket-RL as a practical pathway to decentralized markets that are economically efficient, socially equitable, and technically robust.

Future work will tighten realism and broaden applicability by incorporating distribution-network limits, dynamic tariffs and carbon signals, and explicit battery-degradation costs; advancing learning with constrained/multi-objective RL and privacy-preserving (federated) training; and improving the LLM critic through calibration, interpretability, and light-weight governance. We also plan to benchmark alternative clearing mechanisms and integrate flexible demand/EV resources, alongside targeted pilots to assess regulatory fit and user acceptance.

## Availability of Data and Materials

The source code for this project is available at: [https://github.com/SHRENIK000/fairmarket-rl-llm-fairness-energy-markets](https://github.com/SHRENIK000/fairmarket-rl-llm-fairness-energy-markets)

**Funding:** The author's work was supported by the University of Michigan-Dearborn's Office of Research "Research Initiation & Development."